\documentclass[%
 reprint,
 amsmath,amssymb,
 aps,
 longbibliography,
pra,
showkeys]{revtex4-2}

\usepackage{amsmath,amssymb}
\usepackage{graphicx}
\usepackage{hyperref}
\usepackage{braket}

\begin{document}
\raggedbottom
\preprint{APS/123-QED}

\title{Conditional squeezing induced by a two-level system: arbitrary-time Magnus coefficients in the quantum Rabi model}

\author{Phoenix M. M. Paing}
\email{phoenix.paing@mail.utoronto.ca}
\author{Daniel F. V. James}%
 \email{dfvj@physics.utoronto.ca}
\affiliation{ 
Department of Physics, University of Toronto, 60 St George St., Toronto ON M5S 1A7
}%
\date{\today}

\begin{abstract}
We present a systematic Magnus expansion treatment of the quantum Rabi model beyond the Rotating Wave Approximation. We show that at the second order of Magnus series, the second-order evolution operator contains a term that induces conditional squeezing of the field mode depending on the state of the atom, in addition to the energy shifts. We analyze the scaling behavior of the conditional squeezing coefficient for $^{87}\mathrm{Rb}$ $5^2S_{1/2}\rightarrow5^2P_{1/2}$ transition line and show that the slow envelope of the squeezing coefficient is maximized at half-detuning cycles, and that it scales with $\frac{4g^2}{\omega_0|\Delta|}$. We also show that the quadrature squeezing angle suggests a possible route towards quantum non-demolition readouts, while further investigation is required for a full first-order suppression. We then connect our work to the well-studied AC-Stark shift and Bloch-Siegert shift using the effective Hamiltonian theory. Finally, we show how the energy shifts and the conditional squeezing arise, as a whole $\mathrm{SU}(1,1)$ algebra, and how they can be disentangled as individual unitary evolutions.
\end{abstract}

\maketitle

\section{Introduction}
Squeezed states of light are among the central nonclassical states in quantum optics. They were first introduced as a way to reduce quantum fluctuations below the shot-noise limit and have since become important in precision measurement, interferometry, quantum information, and gravitational-wave detection \cite{cave, LIGO, shot_noise}. In the context of light--matter interaction, squeezing in the Jaynes-Cummings and quantum Rabi models has also been studied for several decades \cite{Scully_Zubairy_1997}. In particular, the possibility of field squeezing generated through interaction with a two-level atom is discussed in the standard quantum-optics literature and in earlier analytical and numerical studies of the Jaynes--Cummings model \cite{kuklinski, Xie1996}. Thus, the appearance of field squeezing in atom-field models is not, by itself, new.
\par
A more recent research direction concerns conditional or spin-dependent squeezing, in which the squeezing angle applied to a field mode depends on the state of a two-level system. Such conditional squeezing has attracted interest in several platforms, including trapped ions \cite{sutherland}, dispersive Jaynes-Cummings-type systems \cite{Ayyash_cond_squeezing}, and qubit--oscillator microwave-cavity architectures \cite{blumenthal, cond_squeeze_1, cond_squeeze_2}. These articles show that spin-dependent Gaussian operations can be useful for quantum control, state preparation, universal continuous-variable quantum computing, and measurement \cite{lloyd_braunstein}. The present work contributes to this growing literature from a different perspective: instead of beginning with an engineered controlled-squeezing protocol, we derive the conditional squeezing term directly from the native quantum Rabi model and analyze its temporal behavior and the parameter dependence of the conditional squeezing magnitude and angle, suggesting natural routes towards Quantum Non-Demolition (QND)-style readouts.
\par
The rotating wave approximation (RWA) is one of the most widely used approximations in quantum optics. It removes the counterrotating terms from the quantum Rabi Hamiltonian and leads to the Jaynes-Cummings model \cite{JC_OG}. Nevertheless, the counterrotating terms are known to produce measurable corrections, such as the Bloch-Siegert shift \cite{bloch}. The validity of the RWA is analyzed in Burgarth's paper \cite{Burgarth_2024}. Care must always be taken when applying the RWA since premature application can result in local-causality violations \cite{malonni_knight, malonni_james, ClerkSipe1998, sabin_causality}. In this work, we keep the counterrotating terms and ask what physical structures appear order by order when the full quantum Rabi interaction is treated perturbatively using the Magnus expansion. By doing so, we see that conditional squeezing arises as a cross-contribution between rotating and counterrotating terms.
\par
In addition to perturbative methods, there exist effective Hamiltonian methods, including time-averaging \cite{Gamel_2010, James_jerke}, and coarse-graining in time \cite{Macr__2023}. By taking the time derivative of the Magnus generator (see section \ref{IIA}), we construct an effective Hamiltonian associated with the arbitrary-time dynamics. This provides a direct way to recover familiar shift terms, including the AC-Stark and Bloch-Siegert shifts, from the Magnus-expansion treatment while retaining the full time dependence of the coefficients.
\par
The main result of this paper is that the second-order unitary evolution operator contains both familiar energy-shift contributions and a conditional two-photon squeezing term. We derive closed-form expressions for the shift coefficient and the conditional squeezing coefficient ($\zeta(t)$), and analyze its magnitude ($|\zeta(t)|$) and phase ($\arg \zeta(t)$), and the associated quadrature angle. The ($^{87}\mathrm{Rb}$) D1 line is used as a representative example to estimate scales and illustrate the dependence of the squeezing magnitude on detuning and transition frequency. We also show that the shift and squeezing operators close under an ($\mathrm{SU}(1,1)$) algebra, providing a compact algebraic framework for understanding the structure of the second-order dynamics and the relationship between the shift and squeezing contributions.
 \par
 This paper is arranged as follows. In Section \ref{methods}, we present a brief discussion of methods such as the Magnus expansion, effective Hamiltonians, and the quantum Rabi Model. In Section \ref{ME}, we present our results of Magnus evolutions, with each subsection highlighting individual components of the second-order operator. Finally, Section \ref{concl} summarizes our conclusions and highlights future work.

\section{\label{methods}Methods}

\subsection{\label{IIA} The Magnus expansion and the effective Hamiltonians}
The Magnus Expansion was first introduced by Wilhelm Magnus in 1954 \cite{Magnus1954}. The method provides an expansion of the unitary time-evolution operator ($\hat{U}_M(t) = exp(\hat{\Omega}(t))$) in the form of a nested commutator of the Hamiltonian at different times. In the form $\hat{\Omega}(t) = \hat{\Omega}_1(t) + \hat{\Omega}_2(t)+...$, the first and second order representations of $\hat{\Omega}(t)$ is given as follows \cite{Blanes_2009}.
\begin{align}
    &\hat{\Omega}_1(t) = -\frac{i}{\hbar}\int_0^t \hat{H} (t_1) dt_1\\
    &\hat{\Omega}_2(t) = \frac{1}{2}(-\frac{i}{\hbar})^2\int_0^tdt_1\int_0^{t_1} dt_2 [\hat{H}(t_1), \hat{H}(t_2)]
\end{align}
There has been work where one goes beyond the second order in the Magnus Expansion \cite{Begzjav_2017, LU2005247}. For the scope of this research, the second order is adequate to show beyond the RWA phenomenon. One of the main advantages of the Magnus expansion is that the evolution operator remains unitary at every truncation order. This makes it possible to identify the operator structures generated at each perturbative order and to interpret them as effective unitary processes.
\par
The Magnus Expansion converges if the following condition is satisfied \cite{Blanes_2009}.
\begin{equation}
    \frac{1}{\hbar}\int_0^t ||\hat{H}(t')||dt' < \pi
\end{equation}
Here, $||\hat{H}(t')||$ refers to the operator norm of the Hamiltonian. For the quantum Rabi model in a finite photon-number space truncated by cutoff photon number N, the expansion converges if $t<\frac{\pi}{2g\sqrt{N}}$.
\par
In addition to the Magnus expansion, there exist approximate methods of effective Hamiltonian modelling where one models the unitary evolution operator as follows by defining an effective Hamiltonian ($\hat{H}^{eff}(t)$).
\begin{equation}
    \hat{U}^{eff}(t) = \exp(-\frac{i}{\hbar}\int_0^t \hat{H}^{eff}(t')dt')
\end{equation}
Effective Hamiltonian theories exist in the form of time-averaging \cite{James_jerke, Gamel_2010}, coarse-graining \cite{Macr__2023}, and so forth. For the Magnus expansion, one can derive the effective Hamiltonian from the Magnus unitary evolution operator as follows.
\begin{align}
   &\hat{U}^M(t) = \hat{U}^{eff}(t)  \nonumber\\
   &\hat{\Omega}_n(t) = -\frac{i}{\hbar}\int_0^t \hat{H}^{eff}_n(t')dt' \nonumber\\
   &\hat{H}^{eff}_n(t) = i\hbar\frac{d\hat{\Omega}_n(t)}{dt}
   \label{eff}
\end{align}
Similar to the Magnus evolution operator, the index n denotes the approximation order in the effective Hamiltonian. From this point forward, we are not interested in $\hat{U}^{eff}$ and hence, we will drop the index ``M" in the Magnus unitary evolution operator. $\hat{U}(t)$ will now refer to the Magnus unitary evolution operator.
\subsection{\label{JC_ROT} The quantum Rabi model in the rotating frame}

The Hamiltonian described by the quantum Rabi model for a two-level atom can be broken down into the field part $\hat{H}_F(t)$, the atom part $\hat{H}_{at}(t)$, and the interaction part $\hat{H}_I (t)$ as follows.
\begin{align}
    \hat{H}(t) & =  \hat{H_F}(t) + \hat{H}_{at} (t) + \hat{H}_I(t) \\ & = \hbar\omega\hat{a}^\dag \hat{a} + \frac{\hbar \omega_0}{2}\hat{\sigma}_z + \hbar g(i\hat{a}^\dag \hat{\sigma}_+ + i\hat{a}^\dag \hat{\sigma}_- + h.a.)
\end{align}
Here $\hbar$ represents the Planck's constant divided by $2\pi$, i the imaginary number, g the coupling strength between the atom and the field, $\omega$ the frequency of the cavity field mode, $\omega_0$ the transition frequency of the two-level system, $\hat{a}^\dag$ and $\hat{a}$ the photon creation and annihilation operators f the field, $\hat{\sigma}_z$ the Pauli-z operator of the atom, and $\hat{\sigma}_+$ and $\hat{\sigma}_-$ the raising and lowering operator in the atomic Hilbert space. The term $h.a.$ represents the Hermitian adjoint of the operator in brackets. Defining the ``rotated" Hamiltonian as in equation \ref{rot}, one can write down the Schrödinger equation in the ``rotated frame" with a transformed Hamiltonian as follows \cite{gerry2004introductory}.
\begin{equation} 
\hat{H}_{rot}(t) = \hat{U}^\dag (t)\hat{H}(t)\hat{U}(t)-i\hbar \hat{U}^\dag (t) \frac{d\hat{U}(t)}{dt}
\label{rot}
\end{equation}
The Hamiltonian in the  ``rotated" frame is as follows.
\begin{align}
\begin{split}
    \hat{H}'(t) =  \hbar g (&i\hat{a}^\dag \hat{\sigma}_- e^{i\Delta t} -  i\hat{a} \hat{\sigma}_+ e^{-i\Delta t} +\\ &i\hat{a}^\dag \hat{\sigma}_+ e^{i\Sigma t} - i\hat{a} \hat{\sigma}_- e^{-i\Sigma t})
\end{split}
\label{H'}
\end{align}
Here, we defined detuning to be $\Delta = \omega - \omega_0$ and the sum-frequency $\Sigma = \omega + \omega_0$. Notice the terms oscillating at frequency $\Sigma$. Such terms do not contribute much to system dynamics. Physically, the term oscillating at $\Sigma$ destroys a photon and lowers the atom from excited to ground state, while the term oscillating at $-\Sigma$ creates a photon and excites the atom. Under short timescales, the energy-time uncertainty relations allow rapid fluctuations in system dynamics. Many theoretical derivations and experimental applications have neglected such highly oscillating terms by making a so-called ``Rotating Wave Approximation (RWA)" \cite{Rabi1937}. Time-averaging effective Hamiltonian theories have also made similar approximations to the RWA by neglecting higher-order fluctuations by taking ensemble averages of the dynamics \cite{James_jerke}. In this work, we show how going beyond the RWA not only yields the well-known Bloch-Siegert shift but also exhibits conditional squeezing behavior as a cross-contribution between the RWA and counter-RWA terms.

\section{\label{ME}Magnus Expansion on the Rotated Hamiltonian}
\subsection{The First Order Magnus Expansion}
Recall that the first-order Magnus Expansion for the time-evolution operator is as follows.
\begin{equation}
\hat{\Omega}_1(t) = -\frac{i}{\hbar}\int_0^t \hat{H} (t_1) dt_1 
\label{first_order}
\end{equation}
Substituting equation \ref{H'} and integrating over time, we get the first-order time evolution operator as follows.
\begin{align}
\begin{split}
    \hat{\Omega}_1(t) = &\frac{ig}{\Delta}(\hat{a}^\dag \hat{\sigma}_-(1 - e^{i\Delta t}) + \hat{a}\hat{\sigma}_+(1 - e^{-i\Delta t})) +\\&\frac{ig}{\Sigma}(\hat{a}^\dag \hat{\sigma}_+ (1 - e^{i\Sigma t})+ \hat{a}\hat{\sigma}_- (1- e^{-i\Sigma t}))
\end{split}
\end{align}
The validity of RWA is due to the sum-frequency ($\Sigma$) being significantly larger in near-resonant optical systems. In such scenarios, the second term is much smaller than the first term. However, if one chooses $t=\frac{2n\pi}{|\Delta|}$, the detuning-dependent term goes to zero, and the evolution operator consists only of counter-rotating contributions. We refer to such times as "full-detuning cycles". We will also see another set of significant times where the conditional squeezing and AC-Stark shift are maximized. Such times are $t=\frac{(2n+1)\pi}{|\Delta|}$ and we refer to such times as ``half-detuning cycles"\par
Using equation \ref{eff}, we get the first-order effective Hamiltonian as follows.
\begin{equation}
    \hat{H}^{eff}_1(t) =  \hbar g (i\hat{a}^\dag \hat{\sigma}_- e^{i\Delta t} +i\hat{a}^\dag \hat{\sigma}_+ e^{i\Sigma t} + h.a.)
\end{equation}
Notice that the first-order effective Hamiltonian is just the native quantum Rabi Hamiltonian. Therefore, the following orders of the effective Hamiltonian will serve as corrections or perturbations for the quantum Rabi Hamiltonian as long as $\frac{g}{\Delta}$ and $\frac{g}{\Sigma}$ are small. $\Delta \leq\Sigma$ and hence, the criterion for the convergence of the effective Hamiltonian is $\frac{g}{|\Delta|} \ll 1$. 

\subsection{\label{second} The Second Order Magnus Expansion}
Detailed calculation of the second-order Magnus expansion is given in appendix \ref{appdxA}. The second-order Magnus evolution operator is determined as follows.
\begin{align}
    \hat{\Omega}_2(t) = ig^2(\hat{a}^\dag \hat{a}+\frac{1}{2})f(t)\hat{\sigma}_z+\frac{1}{2}\left(\zeta^*(t)\hat{a}^2-\zeta(t)(\hat{a}^\dag)^2\right)\hat{\sigma}_z
    \label{magnus_2}
\end{align}
The mathematical expressions of the time-dependent coefficients are as follows.
\begin{equation}
    f(t) = \frac{t}{\Delta}-\frac{t}{\Sigma}+\frac{\sin(\Sigma t)}{\Sigma^2}-\frac{\sin(\Delta t)}{\Delta^2}
\end{equation}
\begin{equation}
    \zeta(t) = g^2\frac{\omega_0e^{2i\omega t}-\omega e^{i\Sigma t} + \omega e^{i\Delta t} - \omega_0}{\omega(\omega^2 -\omega_0^2)}
    \label{zeta}
\end{equation}
\par
The second-order evolution can be broken down into two parts: the energy-shifts depending on f(t) and the conditional squeezing that depends on $\zeta(t)$. We devote section \ref{shift} for demonstrating energy shifts using the effective Hamiltonian theory, and section \ref{squeeze} for analyzing the conditional squeezing from a two-level atom. Finally, we show how to disentangle the second-order operator into energy-shift and squeezing pieces using $\mathrm{SU}(1,1)$ algebra in section \ref{SU(1,1)}

\subsubsection{\label{shift} Demonstration of AC-Stark shift, Bloch-Siegert shift, and the zero-point energy shift using effective Hamiltonian}

Taking the derivative of the f(t) dependent part of the second-order Magnus operator, we get the effective Hamiltonian governing the energy shifts as follows. 
\begin{align}
    &\{\hat{H}^{eff}_2(t)\}_{shift}= -\hbar g^2(\hat{a}^\dag \hat{a} +\frac{1}{2})\hat{\sigma}_z\frac{df}{dt}\\
    &\frac{df}{dt}=\frac{1}{\Delta}(1-\cos(\Delta t))-\frac{1}{\Sigma}(1-\cos(\Sigma t))
\end{align}
We see that the second-order term is indeed an energy shift term. The term proportional by the number operator $\hat{a}^\dag \hat{a}$ corresponds to the photon-induced energy shifts: the AC-Stark shift that scales with $\frac{1}{\Delta}$, and the Bloch-Siegert shift that scales with $\frac{1}{\Sigma}$. Quantitatively, the AC-Stark shift($\alpha_{AC}(t)$), the Bloch-Siegert shift($\alpha_{BS}(t)$), and the total shift($\alpha(t)$) are expressed as follows.
\begin{align}
    &\alpha_{AC}(t)= \frac{-\hbar g^2N}{\Delta}(1-\cos(\Delta t)) \label{AC}\\
    &\alpha_{BS}(t)= \frac{\hbar g^2N}{\Sigma}(1-\cos(\Sigma t)) \label{BS}\\
    & \alpha(t) = \frac{\hbar g^2N}{\Sigma}(1-\cos(\Sigma t)) - \frac{\hbar g^2N}{\Delta}(1-\cos(\Delta t)) \label{total}
\end{align}
Here, N corresponds to the photon number of the mode. Note that on cycle average, $\bar{\alpha} = \frac{1}{\Sigma}-\frac{1}{\Delta}$ as expected.\par
\begin{figure}[h]
\centering
\includegraphics[width = 0.8\linewidth]{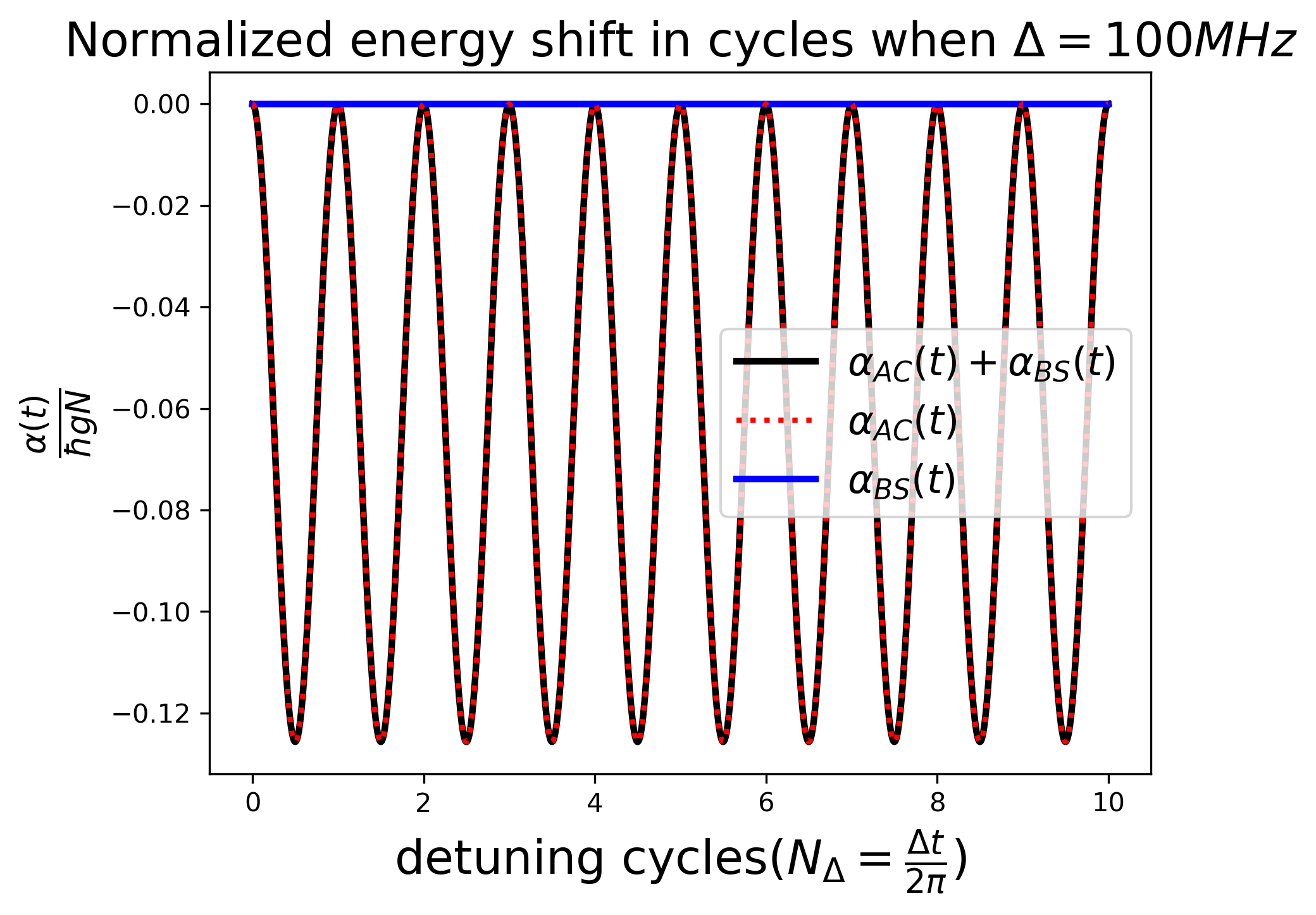}
\caption{This figure shows how the normalized total energy scales with time for $\Delta = 100MHz$. The x-axis represents times in terms of detuning cycles. The y-axis represents the energy shift coefficient normalized by $\hbar g$ and photon number. The similarity of the dotted and solid plots show that the energy-shifts are mainly dominated by the AC-Stark shift.}
\label{fig1}
\end{figure}

The plots in this article are for $^{87}\mathrm{Rb}$ $5^2S_{1/2}\rightarrow5^2P_{1/2}$ line data using $g=2\pi\times1MHz$ \cite{SteckRb87}. Figure \ref{fig1} shows the behavior of energy shifts vs time plot in cycles when the detuning is 100 MHz. 
\par
\begin{figure}[h]
\centering
\includegraphics[width = 0.8\linewidth]{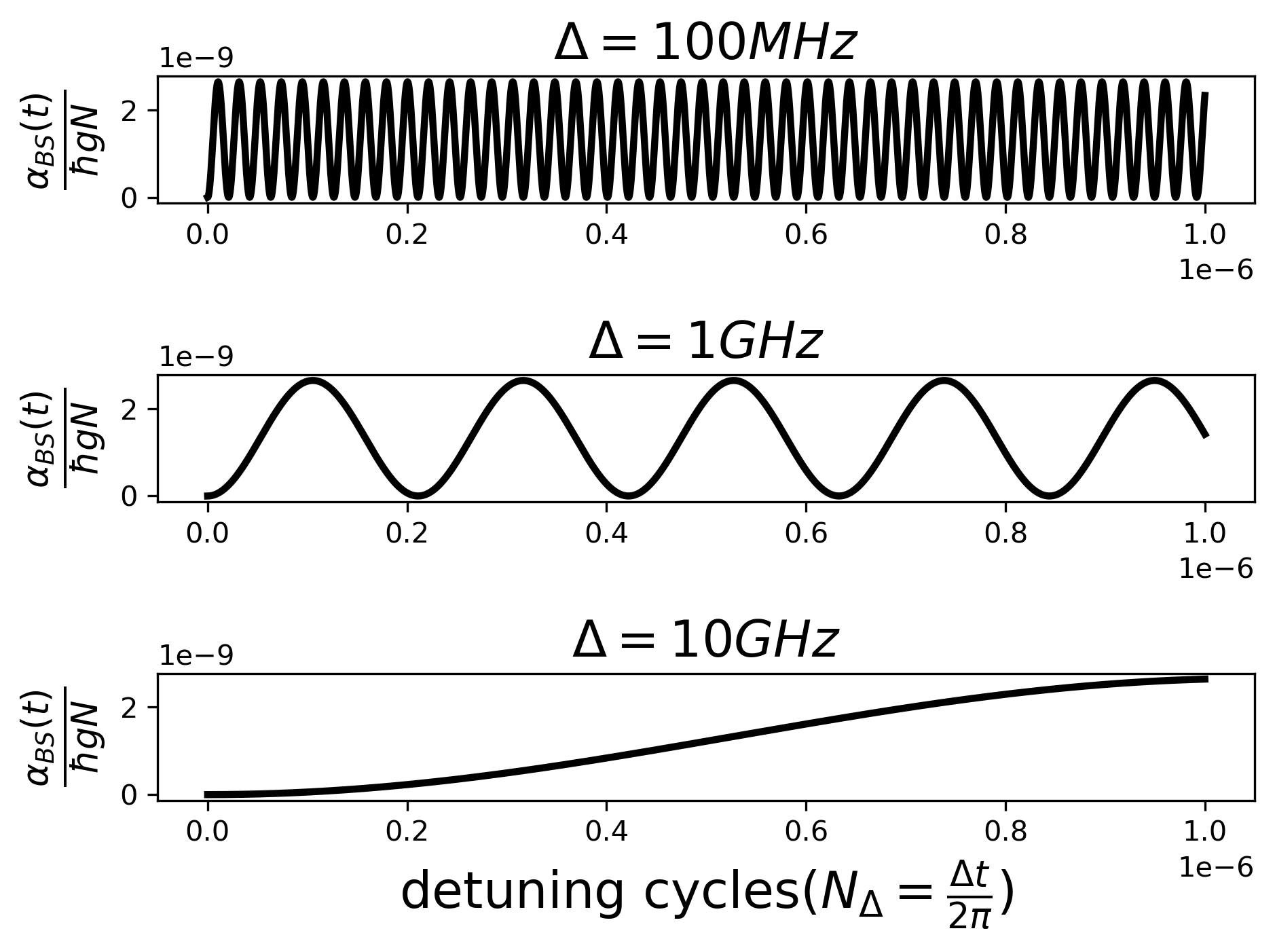}
\caption{This figure shows how the normalized magnitude and oscillation of the Bloch-Siegert shift with respect to microdetuning cycles.}
\label{fig2}
\end{figure}

We can see that the Bloch-Siegert shift acts as a minor correction to the dominant AC-Stark shift. Figure \ref{fig2} shows only the Bloch-Siegert shift contribution at different detunings. Since the Bloch-Siegert shift oscillates rapidly over detuning cycles, the x-axis is scaled into microcycles. The slowing down of the shift with higher detuning is merely because of the frequency mismatch between $\Sigma$ and $\Delta$. The oscillation speed of Bloch-Siegert shift in lab frame has no dependence on detuning.
\par
\begin{figure}[h]
\centering
\includegraphics[width = 0.8\linewidth]{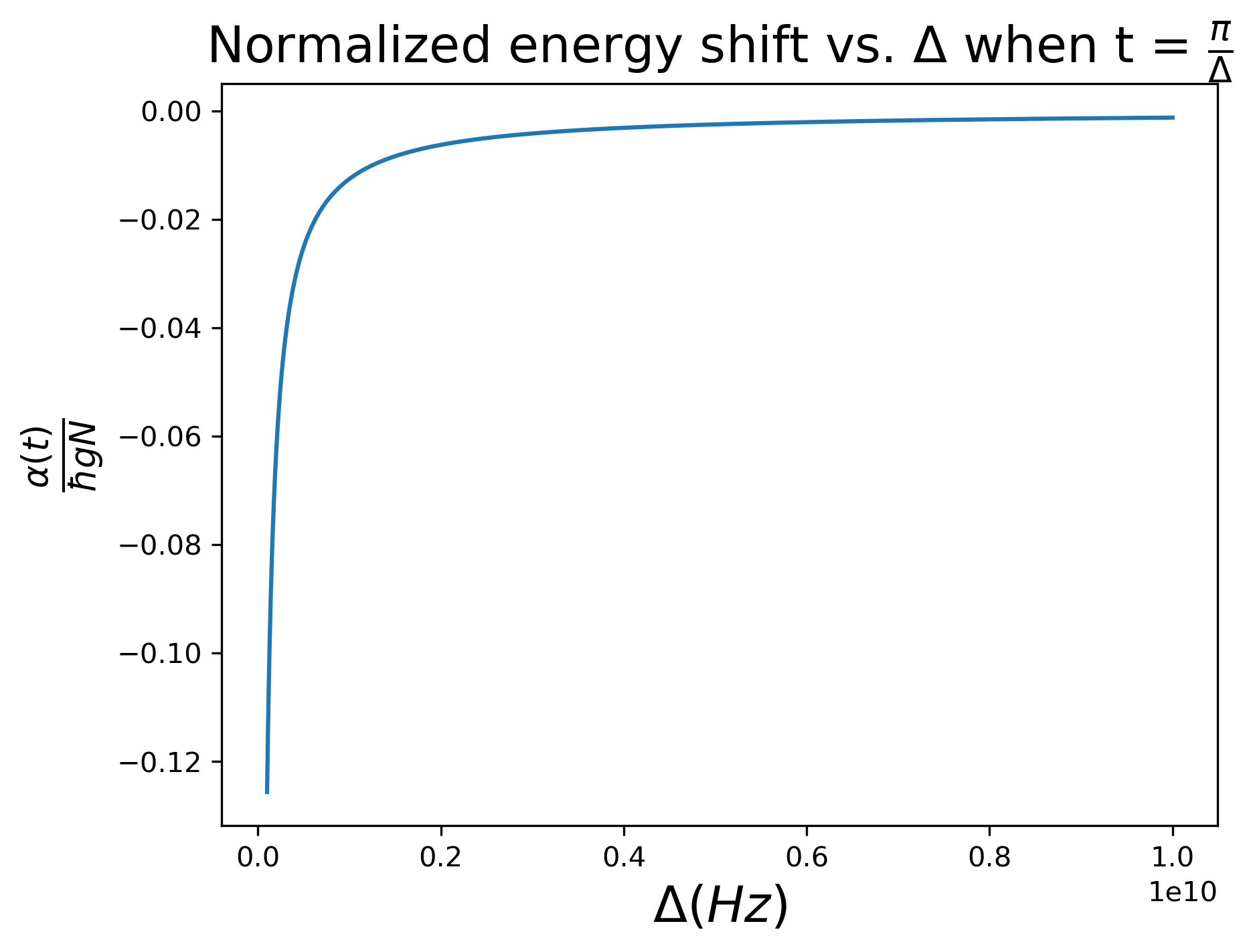}
\caption{This figure shows how the normalized energy shift vs. detuning plot at half detuning cycles.}
\label{fig3}
\end{figure}

Noting that the absolute value of the AC-Stark shift peaks at half-cycles, we plot Figure \ref{fig3} to show the envelope of $\alpha(t)$ as a function of detuning at half-cycles. We see a clean $\frac{1}{\Delta}$ behavior of the energy shift as expected, as the Bloch-Siegert contribution is small. It is also important to note that the AC-Stark shift vanishes at full-detuning cycles, while the Bloch-Siegert shift remains oscillatory. 
\par
The appearance of the 1/2 term in the second-order effective Hamiltonian is noteworthy. Although the bare quantum Rabi Hamiltonian is normally written with the constant zero-point energy of the oscillator removed, the second-order Magnus generator contains the combination $\hat{a}^\dag \hat{a} + 1/2$. This term arises from the noncommutativity of the field operators, $\hat{a}\hat{a}^\dag = \hat{a}^\dag \hat{a} +1$. The appearance of zero-point-energy fluctuation on a model with zero-point-energy omission is an interesting behavior that may require further investigation.

\subsubsection{\label{squeeze} Conditional squeezing from a two-level atom}

The conditional squeezing coefficient in equation \ref{zeta} can be rewritten as follows.
\begin{equation}
    \zeta(t)=\frac{2ig^2e^{i\omega t}}{\omega(\omega^2-\omega_0^2)}(\omega_0 \sin(\omega t)-\omega \sin(\omega_0t))
\end{equation}
Therefore, the fluctuations in the quadrature of the squeezing axes, neglecting the $\mathrm{SU}(1,1)$ disentangling corrections (see section \ref{SU(1,1)}), are as follows.
\begin{align}
    &\langle (\Delta \hat{X}_1)^2\rangle = \frac{1}{4}e^{-2|\zeta(t)|} \nonumber\\
    &\langle (\Delta \hat{X}_2)^2\rangle = \frac{1}{4}e^{2|\zeta(t)|} \nonumber
\end{align}
where,
\begin{equation}
    |\zeta(t)| = \left|\frac{2g^2}{\omega(\omega^2-\omega_0^2)}(\omega\sin(\omega_0 t)-\omega_0\sin(\omega t))\right|
\end{equation}
The squeezing quadrature operator $\hat{X}_1$ is defined on the $\frac{\arg(\zeta(t))}{2}$ axis as the squeezing quadrature, and $\hat{X}_2$ on the $\frac{\arg(\zeta(t))}{2} + \frac{\pi}{2}$ axis as the anti-squeezing quadrature. To calculate the argument of the squeezing coefficient, we first define $C(t)=\frac{2g^2}{\omega(\omega^2-\omega_0^2)}(\omega_0\sin(\omega t)-\omega\sin(\omega_0t))$, which is completely real. Then $\zeta(t) = iC(t)e^{i\omega t} $. Thus the argument is $\pi/2 + \omega t$ if $C(t) >0$, and $3\pi/2 + \omega t$ if $C(t) < 0$. Notice the squeezing angle is undefined when $C(t) = 0$, and this is due to the squeezing coefficient being zero. Moreover, since the squeezing depends on $\hat{\sigma}_z$ operator, the squeezing angle is modified as follows. 
\begin{align}
    &If\  |1\rangle\  and \ C(t) > 0 \rightarrow \arg(\zeta(t)) = \omega t + \frac{\pi}{2}\nonumber \\
    &If\  |1\rangle\  and \ C(t) < 0 \rightarrow \arg(\zeta(t)) = \omega t + \frac{3\pi}{2}\nonumber \\
    &If\  |0\rangle\  and \ C(t) > 0 \rightarrow \arg(\zeta(t)) = \omega t + \frac{3\pi}{2}\nonumber \\
    &If\  |0\rangle\  and \ C(t) < 0 \rightarrow \arg(\zeta(t)) = \omega t + \frac{\pi}{2}
\end{align}
Here, $|0\rangle$ or $|1\rangle$ corresponds to the initial state of the atom being excited or ground state. Notice that the squeezing angle has explicit dependence on $\omega$ but not $\omega_0$. We will see in figure \ref{fig5} that the envelope of squeezing magnitude depends explicitly on $\omega_0$ but not $\omega$. Unlike the energy shifts which are explicit functions of $\Delta$ and $\Sigma$, conditional squeezing coefficients depend on frequencies $\omega$ and $\omega_0$.
\par
The dependence of the squeezing angle on the laser frequency suggests that the squeezed quadrature axis can be controlled through laser parameters. Since the two atomic states generate squeezing along orthogonal quadrature axes, one can use homodyne detection on the outgoing field for QND-style measurements. 
\par
\begin{figure}[h]
\centering
\includegraphics[width = 0.8\linewidth]{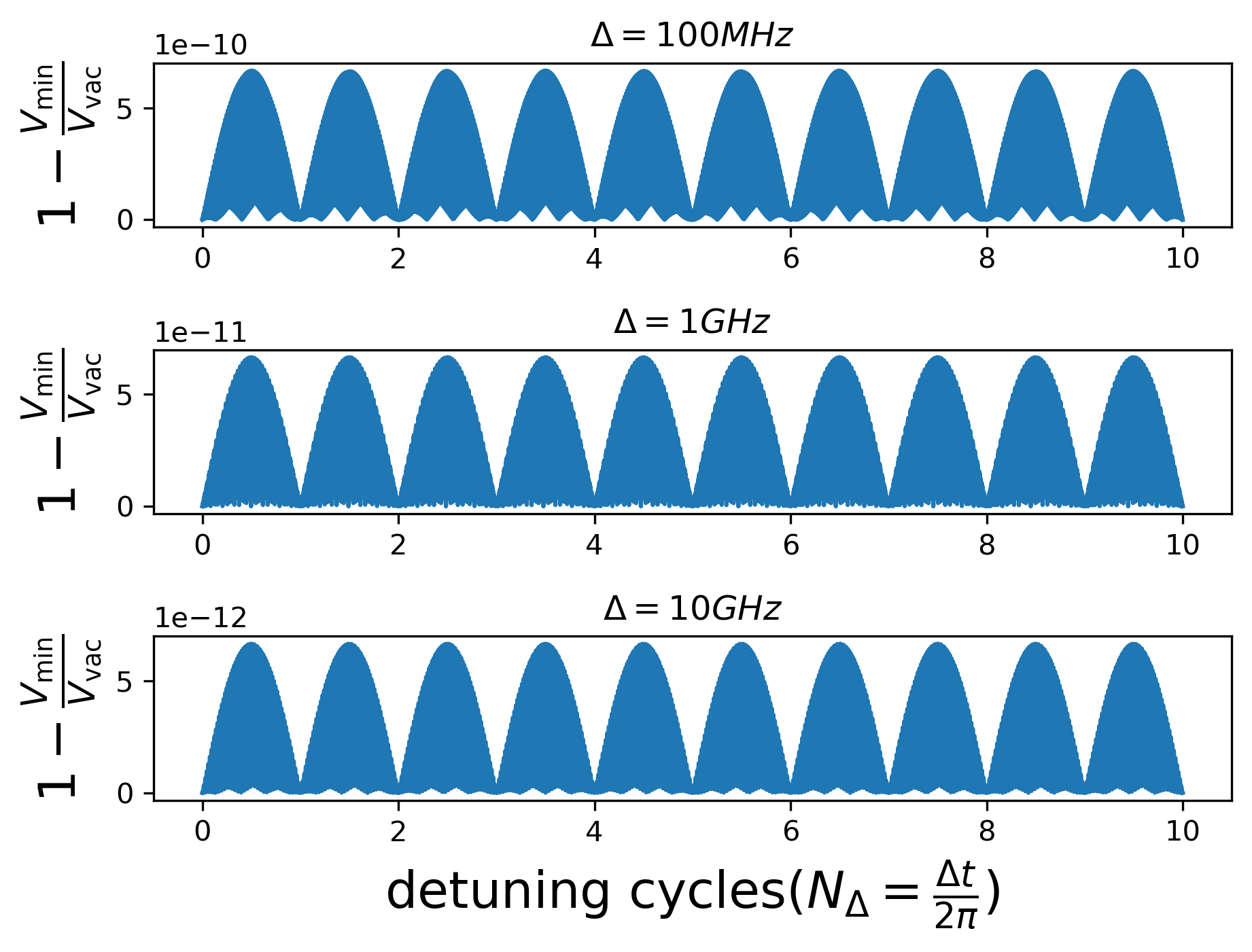}
\caption{This figure shows the squeezing behavior in detuning cycles for different detuning values. The y-axis shows the fractional reduction of the minimum quadrature variance below vacuum. The oscillatory behavior reflects the bounded, nonsecular nature of the squeezing coefficient in the native Rabi model, while the decreasing vertical scale shows the approximate $\frac{1}{|\Delta|}$ suppression of the squeezing envelope at larger detuning.}
\label{fig4}
\end{figure}
Figure \ref{fig4} demonstrates the squeezing magnitude vs time plot for different detuning values. $V_{vac}$ corresponds to the minimum fluctuation of the vacuum, which is $\frac{1}{4}$. $V_{min}$ corresponds to the minimum fluctuation of the squeezed field, or in other words, $\langle (\Delta X_1)^2\rangle$. We see that the oscillation frequency of the plots has no significant dependence on detuning, while the amplitudes increase with lower detuning. Since the squeezing magnitude maximizes at half-detuning cycles, we plot Figure \ref{fig5} to show the scaling of the fractional squeezing magnitude with respect to detuning. Although the squeezing magnitude is small compared with that obtained from conventional nonlinear materials, the explicit $1/\omega_0$ dependence suggests a possible route for enhancing the effect: lower-frequency transitions or engineered effective modes can increase the native conditional squeezing, provided the perturbative condition $g/|\Delta| \ll 1$ remains satisfied. In this sense, the optical $^{87}\mathrm{Rb}$ transition considered here should be viewed as a conservative example rather than an optimized platform. 
\begin{figure}[h]
\centering
\includegraphics[width = 0.8\linewidth]{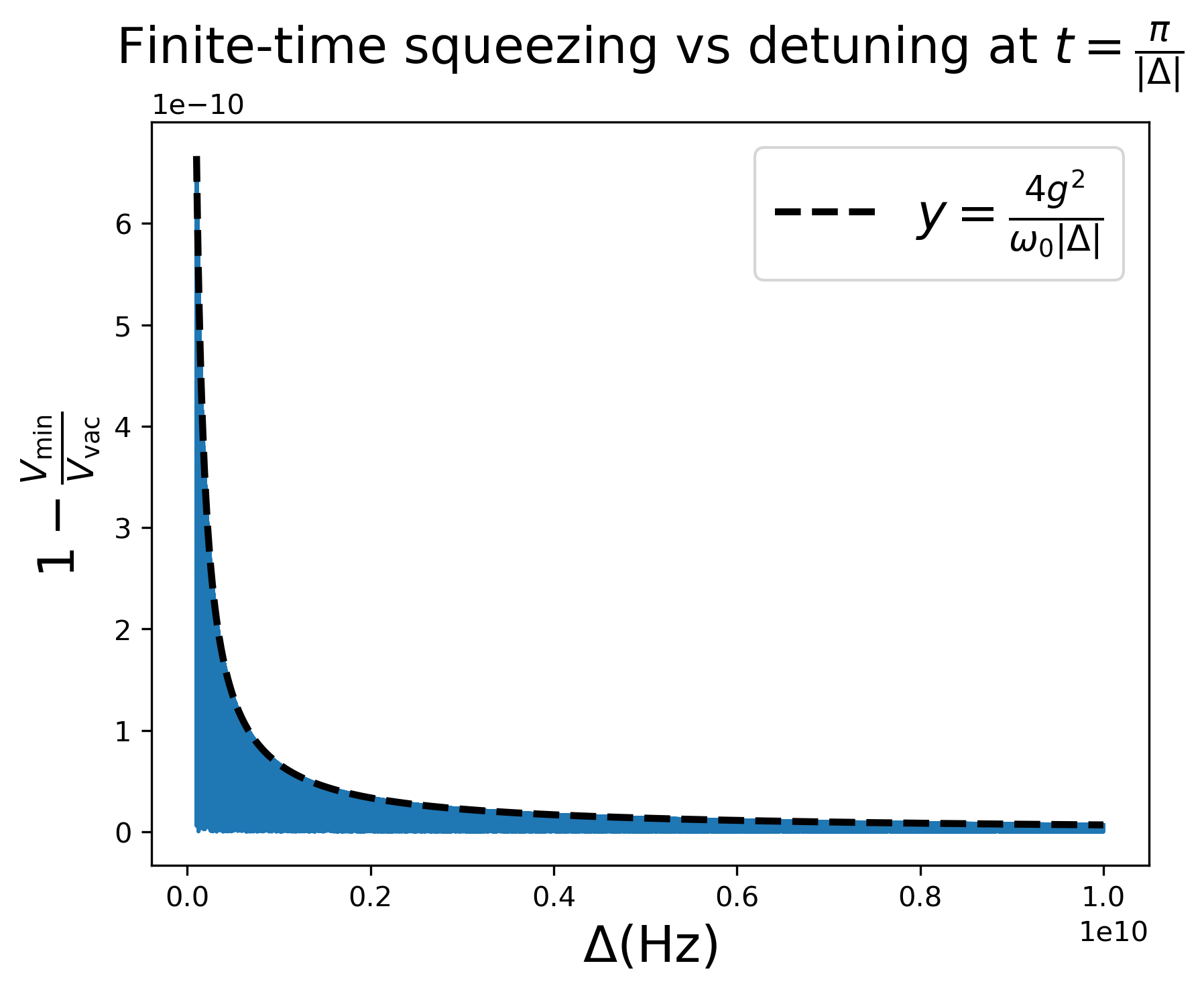}
\caption{This figure shows the squeezing magnitude vs. detuning for a half detuning cycle. It can be seen that the maximum squeezing envelope scales with $\frac{4g^2}{\omega_0|\Delta|}$.}
\label{fig5}
\end{figure}
\par
Moreover, the squeezing vanishes at full detuning cycles and averages to zero over a detuning cycle. Hence, unlike the energy shifts, conditional squeezing does not produce a secular long-time contribution in the quantum Rabi model. Its slow envelope is maximized at the half-detuning cycle, which is also where the instantaneous AC-Stark shift reaches its maximum magnitude. Therefore, the optimal time to observe the finite-time conditional squeezing is at half-detuning cycles.

\subsubsection{\label{SU(1,1)}$\mathrm{SU}(1,1)$ structure and disentanglement}

One normally expects the higher-order terms in a perturbative series to have small corrections to the squeezing coefficient. However, the non-commuting structure embedded in the second order itself induces corrections to the squeezing coefficient without needing to go to higher orders. To demonstrate this, recall the Zassenhaus formula \cite{Zassenhaus}.
\begin{equation}
    e^{\hat{A}+\hat{B}}=e^{\hat{A}}e^{\hat{B}}e^{-\frac{1}{2}[\hat{A},\hat{B}]}e^{\frac{1}{6}(2[\hat{B},[\hat{A},\hat{B}]]+[\hat{A},[\hat{A},\hat{B}]])}+...
\end{equation}
Notice that the commutation relations of the individual field operators in equation \ref{magnus_2} are as follows. 
\begin{align}
    &[\frac{1}{2}(\hat{a}^\dag\hat{a}+\frac{1}{2}), \frac{1}{2}\hat{a}^2] = -\frac{1}{2} \hat{a}^2\\& [\frac{1}{2}(\hat{a}^\dag\hat{a}+\frac{1}{2}), \frac{1}{2}(\hat{a}^\dag)^2] = \frac{1}{2} (\hat{a}^\dag)^2\\&
    [\frac{1}{2}(\hat{a}^\dag)^2, \frac{1}{2}\hat{a}^2] = -(\hat{a}^\dag \hat{a}+\frac{1}{2})
\end{align}
When expanding the exponent using the Zassenhaus formula, one quickly starts noticing that every commutator one evaluates, extra terms of $(\hat{a}^\dag)^2, \hat{a}^2, \hat{a}^\dag \hat{a}+\frac{1}{2}$ are produced. The result is a never-ending loop with every term providing extra squeezing coefficients. However, there is a compact way to evaluate the exponent since the field operators form an $\mathrm{SU}(1,1)$ group.
\par
Three operators $\{\hat{K}_0, \hat{K}_{\pm}\}$ form an $\mathrm{SU}(1,1)$ group if they  obey the following commutation relations \cite{basics_of_SU11}.
\begin{align}
    &[\hat{K}_0, \hat{K}_{\pm}] = \pm \hat{K}_{\pm}\\&
    [\hat{K}_+, \hat{K}_-] = -2\hat{K}_0
\end{align}
Defining $\hat{K_0} = \frac{1}{2}(\hat{a}^\dag\hat{a}+\frac{1}{2})$, $\hat{K}_+ = \frac{1}{2}(\hat{a}^\dag)^2$ and $\hat{K}_- = \frac{1}{2} \hat{a}^2$, and we can see that the operators form an $\mathrm{SU}(1,1)$ group. Then we can rewrite equation \ref{magnus_2} as follows by defining $\alpha(t) = 2ig^2f(t)\hat{\sigma}_z$, and $\beta(t) = -\hat{\sigma}_z \zeta(t)$.
\begin{equation}
    e^{\hat{\Omega}_2(t)} = \exp\left(\alpha(t) \hat{K}_0 + \beta(t)\hat{K}_+ - \beta^*(t)\hat{K}_-\right)
    \label{before_disentangling}
\end{equation}
It is shown in Truax's paper that equation \ref{before_disentangling} can be disentangled into the following form\cite{su11}.
\begin{align}
    e^{\hat{\Omega}_2(t)} =& \exp\left[p_2\hat{K}_+\right]\exp[p_0\hat{K}_0]\exp[p_1\hat{K}_-],
    \label{disentanglement}
\end{align}
The coefficients $p_i(\alpha(t), \beta(t))$ were evaluated in explicit closed forms in Truax's article \cite{su11}. The first commutator between the $\mathrm{SU}(1,1)$ elements already produces terms that scale with $\frac{g^4}{\omega^4}$. Since the correction factors match the scale of fourth-order Magnus operators, evaluating the corrections lies outside the scope of this research.
\par
Note that the Zassenhaus analysis above only applies to the second-order operator. If we were to apply the Zassenhaus formula to $\exp(\hat{\Omega}_1(t))$ instead, we will encounter equations \ref{comm1}--\ref{comm4}. The resulting commutators are $\hat{K}_0$ and $\hat{K}_\pm$, which are elements of the $\mathrm{SU}(1,1)$ group. Thus, the quadratic sector generated at second order is naturally described by an $\mathrm{SU}(1,1)$ structure. Higher-order Magnus terms are therefore expected to generate larger operator algebras, and identifying their structure is an interesting direction for future work.
\par
The full-truncated evolution operator is $\hat{U}(t)\approx \exp(\hat{\Omega}_1(t)+\hat{\Omega}_2(t))$, and $\hat{\Omega}_1(t)$ does not necessarily commute with $\hat{\Omega}_2(t)$. However, the first commutator that they generate ($[\hat{\Omega}_1(t),\hat{\Omega}_2(t)]$) is on the order of $(g/|\Delta|)^3$ and $(g/\Sigma)^3$. Therefore, such terms do not affect the second-order analysis. However, one must include such commutators in the unitary evolution operator if one extends this research to third-order to search for cubic gates. 

\section{\label{concl}Discussion and Conclusion}
In this work, we performed first- and second-order Magnus expansion for the time-evolution operator using the quantum Rabi model in rotated frames. We have seen the non-trivial contributions from the counterrotating terms, such as the Bloch-Siegert shift and the creation of conditionally squeezed light due to the presence of a two-level atom. In this paper, we analyze $|\zeta(t)|$ and $\arg(\zeta(t))$, providing plots for $^{87}\mathrm{Rb}\ D_1$ transition as an example. We find that the squeezing magnitude scales with $\frac{4g^2}{\omega_0|\Delta|}$, suggesting that low-frequency, low-detuning regimes enhance the conditional squeezing behavior. Moreover, we show how the dependence of the squeezing angle on atomic state and natural frequencies suggests a natural route towards quantum state control and Quantum Non-Demolition (QND) measurements. 
\par
However, this QND-style measurement interpretation comes with an important caveat. The full second-order Magnus evolution operator also contains the first-order Magnus terms, which demonstrate raising and lowering operators and can therefore disturb the initialized atomic state. Complete suppression of such first-order effects occurs at full-detuning cycles, but conditional squeezing also vanishes at these times. Thus, the present calculation should not be interpreted as a complete QND measurement protocol. Rather, it identifies the conditional squeezing signature that would need to be isolated in a dispersive or engineered regime where first-order processes are suppressed. One possible direction is to extend this model by adding a qubit drive term to the Hamiltonian as it may allow us to tune the drive frequency resonant to the second-order effects \cite{Ayyash_oscillator}. Such driven extensions may provide a more practical route toward isolating conditional squeezing and non-Gaussian higher-order effects.
\par
We can also notice that the half-cycle squeezing envelope scales as $\frac{4g^2}{\omega_0|\Delta|}$, or equivalently $4\frac{g}{\omega_0}\frac{g}{|\Delta|}$. So an experimentally favorable regime requires suppressing the first-order effects through $\frac{g}{|\Delta|}$ while avoiding an excessively small $\frac{g}{\omega_0}$. For this reason, lower-frequency engineered platforms such as trapped-ion sideband implementations or microwave cavity-QED-systems are favorable for experimental demonstration of conditional squeezing.
\par
Moreover, we calculated the arbitrary-time coefficients of the Bloch-Siegert shift and the AC-Stark shift while highlighting the cosine dependence of the energy shifts using the effective Hamiltonian theorem. We also identified that the AC-Stark shift vanishes on full-detuning cycles, leaving the Bloch-Siegert shift as the sole contribution. At half-detuning cycles, we show that the magnitudes of conditional squeezing and the AC-Stark shift maximize. Finally, we establish a bridge between well-studied energy shifts and the conditional squeezing using $\mathrm{SU}(1,1)$ disentangling.
\par
It is interesting to study the conditional-squeezing phenomena using this approach on multi-level atomic systems. Future research should also investigate whether conditional squeezing exists when an ensemble of atoms is used instead of a single two-level atom. If it exists, it is interesting to determine whether an ensemble of atoms amplifies or attenuates squeezing. Moreover, this research could be directly extended by studying the arbitrary-time coefficients of the third-order Magnus operator, and analyzing the possibility of creating a cubic phase gate to satisfy the Lloyd-Braunstein criterion for UCVQC \cite{lloyd_braunstein, sutherland}. 
\par
Overall, this work provides an arbitrary-time description of conditional squeezing in the quantum Rabi model. Rather than treating the counterrotating terms only as small corrections responsible for frequency shifts, the Magnus expansion reveals that their cross-contributions with the rotating terms also generate a conditional squeezing process. The resulting coefficient ($\zeta(t)$) provides a direct way to quantify both the magnitude and phase of the induced squeezing, while the accompanying shift function (f(t)) connects to the familiar AC-Stark and Bloch-Siegert shifts. This unified description clarifies how energy shifts and conditional squeezing operations emerge from the same beyond-RWA dynamics, and provides a foundation for studying higher-order non-Gaussian effects in light-matter systems.

\begin{acknowledgements}
We thank Mohammad Ayyash for reading the manuscript and providing useful comments. We also appreciate Amar Vutha and all anonymous reviewers of an earlier version for providing useful feedback, which helped us explore new directions and strengthen the manuscript.
\end{acknowledgements}

\section*{Disclosures}

The authors declare no conflicts of interest.

\appendix
\section{\label{appdxA} Derivation of the second-order Magnus Expansion}
Recall that the second-order Magnus Expansion term is given as follows.
\begin{equation}
\hat{\Omega}_2(t) = -\frac{1}{2\hbar^2}\int_0^t dt_1 \int_0^{t_1} dt_2 [\hat{H}' (t_1), \hat{H}' (t_2)]
\label{second_order}
\end{equation}
We get the commutator of the Hamiltonian at two different times as follows.
\begin{align}
    \begin{split}
        -\frac{[\hat{H}' (t_1), \hat{H}' (t_2)]}{\hbar^2 g^2} &= [\hat{a}^\dag \hat{\sigma}_-, \hat{a}\hat{\sigma}_+](-e^{i\Delta(t_1-t_2)} + e^{-i\Delta(t_1-t_2)}) +\\
        & [\hat{a}^\dag \hat{\sigma}_-, \hat{a}^\dag \hat{\sigma}_+](e^{i(\Delta t_1 + \Sigma t_2)} - e^{i(\Sigma t_1 + \Delta t_2)}) +\\
        & [\hat{a} \hat{\sigma}_+, \hat{a} \hat{\sigma}_-](e^{-i(\Delta t_1 + \Sigma t_2)} - e^{-i(\Sigma t_1 + \Delta t_2)}) +\\
        & [\hat{a}^\dag \hat{\sigma}_+, \hat{a}\hat{\sigma}_-](-e^{i\Sigma(t_1-t_2)} + e^{-i\Sigma(t_1-t_2)})
    \end{split}
\label{comm}
\end{align}
Here, we neglect terms involving $[\hat{a}^\dag \hat{\sigma}_-, \hat{a}\hat{\sigma}_-]$ and $[\hat{a}^\dag \hat{\sigma}_+, \hat{a}\hat{\sigma}_+]$ as the commutators evaluate to zero. Since the commutators do not have explicit time dependence, we do the time integrals first and see the contributions of each term present in equation \ref{comm}. The time integrals are as follows.
\begin{align}
    I_1 &= \int_0^{t} dt_1\int_0^{t_1}dt_2 (-e^{i\Delta(t_1-t_2)} + e^{-i\Delta(t_1-t_2)}) \\&= -\frac{2it}{\Delta} +\frac{2i}{\Delta^2}sin(\Delta t) \\
    I_2 &= \int_0^{t} dt_1 \int_0^{t_1} dt_2 (e^{i(\Delta t_1 + \Sigma t_2)} - e^{i(\Sigma t_1 + \Delta t_2)}) \\&=\frac{\omega_0e^{2i\omega t}-\omega e^{i\Sigma t} + \omega e^{i\Delta t} - \omega_0}{\omega(\omega^2 -\omega_0^2)}\\
    I_3 &= \int_0^{t} dt_1 \int_0^{t_1} dt_2 (e^{-i(\Delta t_1 + \Sigma t_2)} - e^{-i(\Sigma t_1 + \Delta t_2)}) \\&= \frac{\omega_0e^{-2i\omega t}-\omega e^{-i\Sigma t} + \omega e^{-i\Delta t} - \omega_0}{(\omega^2 -\omega_0^2)\omega} = I_2^*\\
    I_4 &= \int_0^{t} dt_1\int_0^{t_1}dt_2 (-e^{i\Sigma(t_1-t_2)} + e^{-i\Sigma(t_1-t_2)}) \\&=-\frac{2it}{\Sigma} +\frac{2i}{\Sigma^2}sin(\Sigma t)
    \end{align}
 Note that $I_2$ and $I_3$ being complex conjugates of each other is crucial in yielding the atomic squeezing term later. For notation, call $I_2 = \zeta(t)/g^2$ and $I_3 = \zeta^*(t)/g^2$. \par 
A useful identity for the commutator of products is as follows.
\begin{equation}
    [\hat{A}\hat{B}, \hat{C} \hat{D}] = \hat{A}[\hat{B}, \hat{C}]\hat{D} + \hat{A}\hat{C}[\hat{B}, \hat{D}] + [\hat{A}, \hat{C}]\hat{D}\hat{B} + \hat{C}[\hat{A}, \hat{D}]\hat{B}
\label{commutators}
\end{equation}
Using the commutator identity, we get the following commutator relations.
\begin{align}
    & [\hat{a}^\dag \hat{\sigma}_-, \hat{a}\hat{\sigma}_+] = -\hat{a}^\dag \hat{a} \hat{\sigma}_z - \hat{\sigma}_+\hat{\sigma}_- = -\hat{a}^\dag \hat{a} \hat{\sigma}_z - \ket{1}\bra{1} \label{comm1}\\
    & [\hat{a}^\dag \hat{\sigma}_-, \hat{a}^\dag \hat{\sigma}_+] = -(\hat{a}^\dag)^2 \hat{\sigma}_z \label{comm2}\\
    & [\hat{a} \hat{\sigma}_+, \hat{a}\hat{\sigma}_-] = \hat{a}^2 \hat{\sigma}_z \label{comm3}\\
    & [\hat{a}^\dag \hat{\sigma}_+, \hat{a}\hat{\sigma}_-] = \hat{a}^\dag \hat{a} \hat{\sigma}_z - \hat{\sigma}_-\hat{\sigma}_+ = \hat{a}^\dag \hat{a} \hat{\sigma}_z - \ket{0}\bra{0} \label{comm4}
\end{align}
Here, we define $\ket{1}$ to be the excited state of the atom and $\ket{0}$ its ground state. Combining all terms, we get the full second-order time-evolution operator as follows.
\begin{align}
\begin{split}
    \hat{\Omega}_2(t) =& ig^2(\frac{t}{\Delta}-\frac{\sin(\Delta t)}{\Delta^2})(\hat{a}^\dag \hat{a} \hat{\sigma}_z + \ket{1}\bra{1}) + \\&ig^2(\frac{t}{\Sigma}-\frac{\sin(\Sigma t)}{\Sigma^2})(-\hat{a}^\dag \hat{a} \hat{\sigma}_z + \ket{0}\bra{0} ) +\\
    &\frac{1}{2}\left(\zeta^*(t)\hat{a}^2-\zeta(t)(\hat{a}^\dag)^2\right)\hat{\sigma}_z
\end{split}
\label{2ndorder}
\end{align}
In equation \ref{2ndorder}, we can rewrite $\ket{1}\bra{1}$ as $\frac{1}{2}(\hat{I}_{at}+\hat{\sigma}_z)$ and $\ket{0}\bra{0}$ as $\frac{1}{2} (\hat{I}_{at}-\hat{\sigma}_z)$ where $\hat{I}_{at}$ is the identity operator in the Hilbert Space of the two-level atom. Noting that the identity operator commutes with every other operator, we can ignore the contribution of the term associated with the identity operator, as it only adds a global phase to the system. Hence, we rewrite equation \ref{2ndorder} as follows.
\begin{align}
    \hat{\Omega}_2(t) = ig^2(\hat{a}^\dag \hat{a}+\frac{1}{2})f(t)\hat{\sigma}_z+\frac{1}{2}\left(\zeta^*(t)\hat{a}^2-\zeta(t)(\hat{a}^\dag)^2\right)\hat{\sigma}_z
\end{align}
where $f(t)$ represents the time-dependent shift coefficient and $\zeta(t)$ the conditional squeezing coefficient.
\begin{equation}
    f(t) = \frac{t}{\Delta}-\frac{t}{\Sigma}+\frac{\sin(\Sigma t)}{\Sigma^2}-\frac{\sin(\Delta t)}{\Delta^2}
\end{equation}
\begin{equation}
    \zeta(t) = g^2\frac{\omega_0e^{2i\omega t}-\omega e^{i\Sigma t} + \omega e^{i\Delta t} - \omega_0}{\omega(\omega^2 -\omega_0^2)}
\end{equation}

\bibliographystyle{apsrev4-2-titleparen}
\bibliography{QND_arxiv}

\end{document}